\documentclass[aps,prc,twocolumn,superscriptaddress,showpacs]{revtex4-2}
\usepackage[utf8]{inputenc}
\usepackage{graphicx}% Include figure files
\graphicspath{ {./images/} }
\usepackage{amssymb}
\usepackage{xcolor}
\RequirePackage{amsmath}
\RequirePackage{amsfonts}
\RequirePackage{graphicx}
\usepackage{fontenc}
\usepackage{longtable}
\usepackage{dcolumn}
\usepackage{svrsymbols}
\usepackage{array}
\newcolumntype{P}[1]{>{\centering\arraybackslash}p{#1}}
\begin{document}
%\title{Exploring the origin of the reactor antineutrino anomaly: high-resolution $\beta^{-}$-decay study of $^{92}$Rb}
%\title{$\beta^{-}$ decay of $^{92}$Rb and the population of the Pygmy Dipole Resonance of $^{92}$Sr}
\title{First observation of the fine structure of the Pygmy Dipole Resonance in a nucleus away from stability via the $\beta^-$ decay of $^{92}$Rb to $^{92}$Sr}
\author{P.~Spagnoletti}
\altaffiliation{Corresponding author}
\altaffiliation{Present Address:~Department of Physics, University of Liverpool, Liverpool L69 7ZE, United Kingdom}
\affiliation{Department of Chemistry, Simon Fraser University, Burnaby, British Columbia V5A 1S6, Canada}
\author{M.~Ramalho}
\altaffiliation{School of Physics, Engineering and Technology, University of York, Heslington, York YO10 5DD, United Kingdom}
\author{S.~Tähtelä}
\author{E. Kauppinen}
\affiliation{Department of Physics, University of Jyv\"askyl\"a, P.O. Box 35, FI-40014, Jyv\"askyl\"a, Finland}
\author{J.~Suhonen}
\affiliation{Department of Physics, University of Jyv\"askyl\"a, P.O. Box 35, FI-40014, Jyv\"askyl\"a, Finland}
\affiliation{International Centre for Advanced Training and Research in Physics (CIFRA), P.O. Box MG12, 077125 Bucharest-Magurele, Romania}
\author{C.~Andreoiu}
\affiliation{Department of Chemistry, Simon Fraser University, Burnaby, British Columbia V5A 1S6, Canada}
\author{M.~Scheck}
%\affiliation{School of Computing, Engineering and Physical Sciences, University of the West of Scotland, Paisley, PA1 2BE, United Kingdom}
%\affiliation{FRG, Abt. RR, Helmholtz-Zentrum Hereon, Max Planck Str. 1, 21502 Geesthacht, Germany}
\affiliation{RR, Forschungsreaktor, Helmholtz Zentrum Hereon, 21502 Geesthacht, Germany}
\author{V.~Vedia}
\altaffiliation{Present Address:~CERN, CH-1211 Geneva, Switzerland}
\affiliation{TRIUMF, 4004 Wesbrook Mall, Vancouver, BC, V6T 2A3, Canada}
\author{Z.~Ahmed}
\affiliation{Department of Physics, University of Guelph, Guelph N1G 2W1 Ontario, Canada}
\author{D.~Annen}
\affiliation{Department of Chemistry, Simon Fraser University, Burnaby, British Columbia V5A 1S6, Canada}
\author{G.C.~Ball}
\affiliation{TRIUMF, 4004 Wesbrook Mall, Vancouver, BC, V6T 2A3, Canada}
\author{G.~Benzoni}
\affiliation{INFN Sezione di Milano, IT-20133 Milano, Italy}
\author{S.S.~Bhattacharjee}
\affiliation{TRIUMF, 4004 Wesbrook Mall, Vancouver, BC, V6T 2A3, Canada}
\author{H.~Bidaman}
\affiliation{Department of Physics, University of Guelph, Guelph N1G 2W1 Ontario, Canada}
\author{V.~Bildstein}
\affiliation{Department of Physics, University of Guelph, Guelph N1G 2W1 Ontario, Canada}
\author{S.~Buck}
\affiliation{Department of Physics, University of Guelph, Guelph N1G 2W1 Ontario, Canada}
\author{R.~Caballero-Folch}
\affiliation{TRIUMF, 4004 Wesbrook Mall, Vancouver, BC, V6T 2A3, Canada}
\author{R.J.~Coleman}
\affiliation{Department of Physics, University of Guelph, Guelph N1G 2W1 Ontario, Canada}
\author{S.~Devinyak}
\affiliation{TRIUMF, 4004 Wesbrook Mall, Vancouver, BC, V6T 2A3, Canada}
\author{I.~Dillmann}
\affiliation{TRIUMF, 4004 Wesbrook Mall, Vancouver, BC, V6T 2A3, Canada}
\affiliation{Department of Physics and Astronomy, University of Victoria, Victoria, British Columbia, Canada}
\author{I.~Djianto}
\affiliation{Department of Chemistry, Simon Fraser University, Burnaby, British Columbia V5A 1S6, Canada}
\author{F.H.~Garcia}
\altaffiliation{Present Address: Department of Chemistry, Simon Fraser University, Burnaby, British Columbia V5A 1S6, Canada}
\affiliation{Lawrence Berkeley National Laboratory, Berkeley, California 94720, USA}
\author{A.B.~Garnsworthy}
\affiliation{TRIUMF, 4004 Wesbrook Mall, Vancouver, BC, V6T 2A3, Canada}
\author{P.E.~Garrett}
\affiliation{Department of Physics, University of Guelph, Guelph N1G 2W1 Ontario, Canada}
\author{B.~Greaves}
\affiliation{Department of Physics, University of Guelph, Guelph N1G 2W1 Ontario, Canada}
\author{C.J.~Griffin}
\affiliation{TRIUMF, 4004 Wesbrook Mall, Vancouver, BC, V6T 2A3, Canada}
\author{G.F.~Grinyer}
\affiliation{Department of Physics, University of Regina, S4S 0A2 Regina, Canada}
\author{E.G.~Fuakye}
\affiliation{Department of Physics, University of Regina, S4S 0A2 Regina, Canada}
\author{G.~Hackman}
\affiliation{Department of Chemistry, Simon Fraser University, Burnaby, British Columbia V5A 1S6, Canada}
\affiliation{TRIUMF, 4004 Wesbrook Mall, Vancouver, BC, V6T 2A3, Canada}
\author{D.~Hymers}
\affiliation{Department of Physics, University of Guelph, Guelph N1G 2W1 Ontario, Canada}
\altaffiliation{Institute for Nuclear Physics, University of Cologne, 50937 Cologne, Germany}
\author{D.~Kalaydjieva}
\affiliation{IRFU/DPhN, CEA Saclay, Universit\'e Paris-Saclay, 91191 Gif-sur-Yvette, France}
%\altaffiliation{Present Address:~Department of Physics, University of Guelph, Guelph N1G 2W1 Ontario, Canada}
\author{R.~Kanungo}
\altaffiliation{Present Address:~TRIUMF, 4004 Wesbrook Mall, Vancouver, BC, V6T 2A3, Canada}
\affiliation{Astronomy and Physics Department, Saint Mary's University, Halifax, Nova Scotia, B3H 3C3, Canada}
\author{K.~Kapoor}
\affiliation{Department of Physics, University of Regina, S4S 0A2 Regina, Canada}
\author{E.~Kasanda}
\affiliation{Department of Physics, University of Guelph, Guelph N1G 2W1 Ontario, Canada}
\author{W.~Korten}
\affiliation{IRFU/DPhN, CEA Saclay, Universit\'e Paris-Saclay, 91191 Gif-sur-Yvette, France}
\author{N.~Marchini}
\affiliation{INFN Sezione di Firenze, IT-50019 Firenze, Italy}
\author{K.~Mashtakov}
\affiliation{Department of Physics, University of Guelph, Guelph N1G 2W1 Ontario, Canada}
\author{A.~Nannini}
\affiliation{INFN Sezione di Firenze, IT-50019 Firenze, Italy}
\author{K.~Ortner}
\altaffiliation{Present Address:~TRIUMF, 4004 Wesbrook Mall, Vancouver, BC, V6T 2A3, Canada}
\affiliation{Department of Chemistry, Simon Fraser University, Burnaby, British Columbia V5A 1S6, Canada}
\author{B.~Olaizola}
\altaffiliation{Present Address:~Instituto de Estructura de la Materia, CSIC, 28006, Madrid, Spain}
\affiliation{CERN, CH-1211 Geneva, Switzerland}
\author{C.~Natzke}
\affiliation{TRIUMF, 4004 Wesbrook Mall, Vancouver, BC, V6T 2A3, Canada}
\affiliation{Department of Physics, Colorado School of Mines, Golden, Colorado 80401, USA}
\author{C.M.~Petrache}
%\affiliation{IRFU/DPhN, CEA Saclay, Universit\'e Paris-Saclay, 91191 Gif-sur-Yvette, France}
\affiliation{Universit\'e Paris-Saclay, CNRS/IN2P3, IJCLab, 91405 Orsay, France}
\author{M.~Polettini}
%\affiliation{INFN Sezione di Firenze, IT-50019 Firenze, Italy}
\affiliation{INFN Sezione di Milano, Milano 20133, Italy}
\affiliation{Università degli Studi di Milano, Milano 20133, Italy}
\author{A.~Radich}
\affiliation{Department of Physics, University of Guelph, Guelph N1G 2W1 Ontario, Canada}
\author{M.~Rocchini}
\affiliation{Department of Physics, University of Guelph, Guelph N1G 2W1 Ontario, Canada}
\affiliation{INFN Sezione di Firenze, IT-50019 Firenze, Italy}
\author{N.~Saei}
\affiliation{Department of Physics, University of Regina, S4S 0A2 Regina, Canada}
\author{M.~Satrazani}
\altaffiliation{Present Address:~KU Leuven, Instituut voor Kern- en Stralingsfysica, 3001 Leuven, Belgium}
\affiliation{Department of Physics, University of Liverpool, Liverpool L69 7ZE, United Kingdom}
\author{M.~Siciliano}
\affiliation{Physics Division, Argonne National Laboratory, Argonne, Illinois 60439, USA}
\author{M.~Singh}
\affiliation{Astronomy and Physics Department, Saint Mary's University, Halifax, Nova Scotia, B3H 3C3, Canada}
\author{C.E.~Svensson}
\affiliation{TRIUMF, 4004 Wesbrook Mall, Vancouver, BC, V6T 2A3, Canada}
\affiliation{Department of Physics, University of Guelph, Guelph N1G 2W1 Ontario, Canada}
\author{D.A.~Torres}
\affiliation{Departamento de Física, Universidad Nacional de Colombia, Bogotá, Colombia}
\author{R.~Umashankar}
\affiliation{TRIUMF, 4004 Wesbrook Mall, Vancouver, BC, V6T 2A3, Canada}
\affiliation{Department of Physics and Astronomy, University of British Columbia, Vancouver, British Columbia, Canada V6T 1Z4}
\author{E.~Wadge}
\affiliation{Department of Chemistry, Simon Fraser University, Burnaby, British Columbia V5A 1S6, Canada}
\author{T.~Zidar}
\affiliation{Department of Physics, University of Guelph, Guelph N1G 2W1 Ontario, Canada}
\author{M.~Zieli\'nska}
\affiliation{IRFU/DPhN, CEA Saclay, Universit\'e Paris-Saclay, 91191 Gif-sur-Yvette, France}

\begin{abstract}
A comprehensive $\gamma$-ray spectroscopy study of excited states in $^{92}$Sr populated via $\beta^-$ decay of $^{92}$Rb ($J^{\pi}=0^-$, $Q_{\beta}=8095(6)$~keV) was performed with the GRIFFIN spectrometer at the Isotope Separator and Accelerator (ISAC) radioactive ion beam facility at TRIUMF.
The high $\gamma$-ray efficiency of the GRIFFIN spectrometer combined with the intense rubidium beams produced allowed for 864 $\gamma$-ray transitions to be placed in the level scheme with 190 excited levels populated, most of them identified for the  first time.
The excitation energies of low-spin states in $^{92}$Sr are well reproduced by large-scale Shell Model calculations up to 5~MeV. 
%Almost all of these levels and $\gamma$-ray transitions have been identified for the first time.
The $\beta$-feeding intensities are in very good agreement with results from a recent study employing Modular Total Absorption Spectroscopy, indicating a significant suppression of the Pandemonium effect.
%The experimental picture reveals an unprecedented detail of energy-dependent log~$ft$ values.
The experimental picture reveals the energy level dependence of log~$ft$ values in unprecedented detail.
These log~$ft$ values are well reproduced by Multiple-Commutator Model calculations that identify the features in the excited levels' wavefunctions that enable the population of high-lying levels with a low effective Q-value that belong to the Pygmy Dipole Resonance.

\end{abstract}
\maketitle
%\section{Introduction\label{sec:int}}
The Pandemonium effect~\cite{HARDY1977307} is the name given to a systematic error when measuring the $\beta$-feeding intensities with high-resolution but low-efficiency $\gamma$-ray detectors.
This effect is particularly pronounced for decays with a large $Q_{\beta}$ value to the daughter nucleus and describes the non-observation or misplacement of $\gamma$-ray transitions within the level scheme due to the finite resolving power of high-resolution $\gamma$-ray spectroscopy. 
This limitation can result in incorrect $\beta$-feeding intensity measurements leading to serious negative outcomes beyond the realm of nuclear structure which is typically the focus of such studies. \\
The Reactor Antineutrino Anomaly (RAA) refers to a $\sim6\%$ deficit of the antineutrino flux measured in separate experiments~\cite{DayaBayFlux,DayBay-2019,RENO-Flux,DoubleChooz2020} when compared to predictions of the Huber-Mueller (HM) model~\cite{Huber,Mueller}, which has prompted a flurry of activity over the past 15 years.
%A major ingredient of this anomaly is that much of the existing $\beta$-decay data which serve as a key input for model predictions of the reactor flux suffer from this Pandemonium effect 
These electron antineutrinos are produced via the $\beta^{-}$~decay of fission products from the reactor and their measured properties serve as a vital component in understanding the RAA.
As such, the Summation Method (SM), which computes the antineutrino energy spectrum produced from individual $\beta$ transitions and sums over all transitions for all of the decaying fission fragments provides an alternative approach to the HM model~\cite{RAA_review}.
The success of this approach relies upon robust measurements of $\beta$-feeding intensities free of Pandemonium of a large number of $\beta$-decaying nuclei that are strongly produced via nuclear fission.
To fulfill this need, Total Absorption Spectroscopy methods have become the established approach to obtain Pandemonium free $\beta$-decay data.
However, while this approach has proven highly effective in revealing the previously underestimated population of high-lying levels via $\beta$-decay~\cite{PhysRevLett.123.022502}, the method itself is a limited probe into the structure of these levels.
The energy dependence upon the probability for competing $\beta$~decays is sensitive to the effective $Q$ value ($Q_{eff}=Q_{\beta}-E_x$) to the power of five, with $E_x$ defined as the excitation energy of the daughter.
This relationship raises the question as to which features of the high-lying levels' wavefunction enables them to compete with decays to lower-lying levels which are significantly energetically favored. \\
The $\beta$~decay of $^{92}$Rb ($T_{1/2}=4.48(3)$~s) possesses a $J^{\pi}=0^{-}$ ground state and $Q_{\beta}=8095(6)$~keV~\cite{DATASHEETSA92}. 
This decay is a main contributor to the high-energy region of the reactor antineutrino spectrum owing to the high $Q_{\beta}$ value and large yield following neutron-induced fission. 
The significance of this decay motivated two independent measurements using Total Absorption Spectroscopy techniques~\cite{TAGS_92Sr,Rasco_92Rb} which revealed a significant increase in $\beta$-feeding to excited levels above 4.5~MeV compared to an earlier measurement using high-resolution spectroscopy~\cite{92Rb_decay_1972} and a more recent study of the $\beta$-spectral shape of this decay~\cite{algora-bssm}. 
What remains unexplored from these studies is an investigation into the structure of these levels and which features of the high-lying levels' wavefunction enables them to compete with decays to lower-lying levels which are significantly energetically favored.\\
The $\beta$~decay of $^{92}$Rb exhibits similar characteristics to the $\beta$~decay of the neighboring nucleus $^{96}$Y ($Q_{\beta}=7109(6)$~keV, $J^{\pi}=0^-$) which is another main contributor to the high-energy region of the reactor antineutrino spectrum.
A recent study of this decay~\cite{mashtakov_96Zr} revealed significant $\beta$-feeding via allowed Gamow-Teller transitions to the resonance-like structure of $1^-$ levels in $^{96}$Zr known as the Pygmy Dipole Resonance (PDR)~\cite{SAVRAN2013210}.
The opportunities to investigate the PDR via $\beta$~decay is limited to decays with the parent nucleus possessing a negative-parity low-spin ground state and sufficient $Q_{\beta}$ to populate levels near the neutron separation energy of the daughter~\cite{Scheck_PDR_PRL}.
Consequently, experimental and theoretical studies of the PDR via $\beta$~decay are scarce~\cite{mashtakov_96Zr,Scheck_PDR_PRL,GOTTARDO2017359,PDR_80Ge,BRACCO_PDR_Review,LANZA2_PDR_Review}.
However, $\beta$~decay offers an alternative approach to the well established $(p,p')$ and $(\gamma,\gamma')$ methods, in particular, the ability to investigate the $\gamma$-ray decay behavior of these levels, provide a complementary insight into the microscopic structure of the wave function of these $1^-$ levels and the opportunity to access the PDR of unstable nuclei such as $^{92}$Sr. 
%The PDR refers to the additional electric-dipole strength situated at the low-energy tail of the isovector Giant Dipole Resonance (GDR). 
%\textcolor{red}{\\ OLD VERSION \\}
%The $\beta$~decay of $^{92}$Rb with its $0^-$ ground state, may populate $0^-$ and $1^-$ levels in $^{92}$Sr via allowed decays while first-forbidden decays populate $0^+$, $1^+$, and $2^+$ levels. 
%The probability for competing $\beta$~decays is sensitive to the effective $Q$ value ($Q_{\beta}-E_x$) to the power of five, with $E_x$ defined as the excitation energy of the daughter.
%This energy dependence raises the question of which features of the high-lying levels' wavefunction enables them to compete with decays to lower-lying levels which are significantly energetically favored. 
This work contributes to the study of the PDR states populated via $\beta^{-}$~decay, provides complementary spectroscopic data to the previous Total Absorption Spectroscopy studies of this decay, which is a key contributor to the high-energy region of the reactor antineutrino energy spectrum.
For this aim, a high-resolution spectroscopy experiment has been performed to reveal a picture of $\beta$~decay with unparalleled detail. \\
Excited states in $^{92}$Sr were populated following $\beta^{-}$~decay of $^{92}$Rb ($J^{\pi}=0^-$, $Q_{\beta}=8095(6)$~keV) at the TRIUMF-ISAC~1 facility~\cite{Dilling2014}.
Radioactive isotopes were produced by bombarding a uranium carbide (UC$_x$) target with 480-MeV protons. % provided by the main 520-MeV cyclotron. %~\cite{Bylinskii2014}. 
The reaction products were ionized using a ruthenium surface ion source and then mass separated to produce a beam of A~=~92 ions.
%passed through the high-resolution mass spectrometer ($M / \delta M~\approx~2000$)~\cite{Bricault2014} to produce a beam of A~=~92 ions.
A 40-keV beam of $^{92}$Rb and $^{92}$Sr ions were implanted into a moving tape situated at the center of the GRIFFIN $\gamma$-ray spectrometer~\cite{GRIFFIN_NIM}.
In this study, GRIFFIN was equipped with 15 Compton-suppressed HPGe clover detectors.
A 10-mm thick Delrin absorber surrounded the implantation chamber at the center of the GRIFFIN array to suppress Bremsstrahlung radiation from high-energy $\beta$ particles.
Decay data were obtained using an implantation time of 45~s, corresponding to approximately ten half-lives of $^{92}$Rb ($T_{1/2}=4.49$~s), and a further 13~s with no implantation. 
%These data were recorded using a custom built digital data acquisition system~\cite{GRFFIN_DAQ}, which collected energy and timing signals in a triggerless mode.
%After this time, the tape was moved and the cycle repeated.
%This tape cycle was employed to suppress contamination from the decay of $^{92}$Sr ($T_{1/2}=2.66$~h) and $^{92}$Y ($T_{1/2}=3.54$~h).
Data were collected at an intensity of 1$\times$10$^6$~pps for 80~minutes, then at 3$\times$10$^6$~pps for six hours.
%Data were collected with a beam intensity of $^{92}$Rb of 1$\times$10$^6$ pps for approximately 80~minutes followed by an increase to 3$\times$10$^6$ pps and data were collected for six hours.
%Later, the intensity of the $^{92}$Rb ions was increased to 3$\times$10$^6$ pps and the data were collected for six hours.
\begin{figure}[t]
\includegraphics[width=\columnwidth]{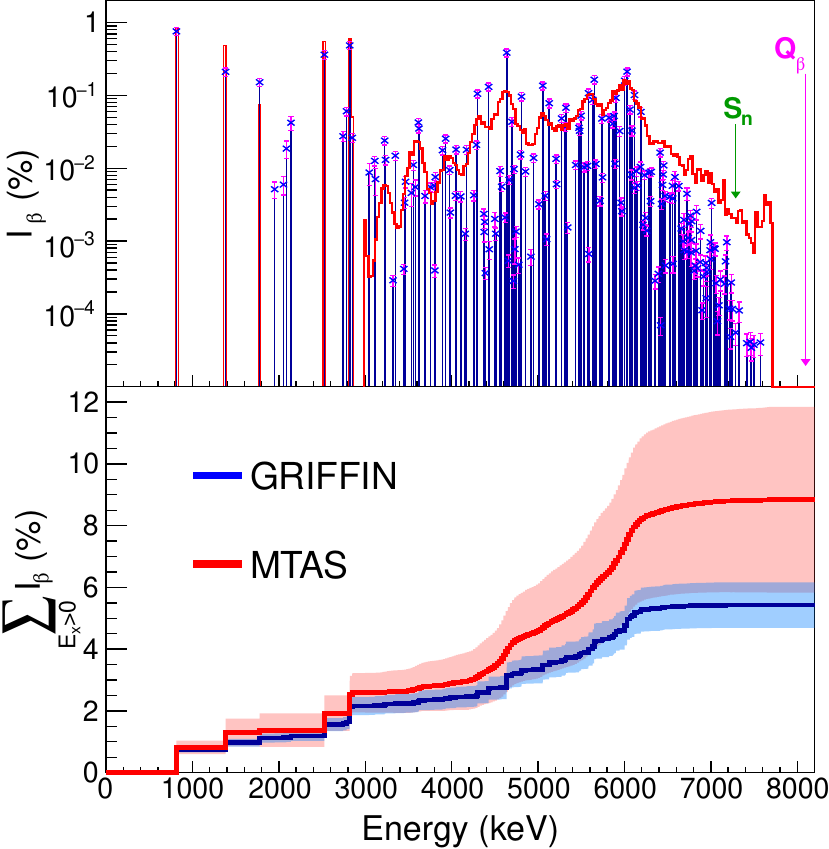}
\caption{Top: $\beta$-feeding probabilities from $^{92}$Rb~$\rightarrow$~$^{92}$Sr decay measured in this work (blue) and from the MTAS measurement~\cite{Rasco_92Rb} (red).
Bottom: Cumulative $\beta$-feeding intensity in $^{92}$Sr excluding the ground-state decay measured in this work (blue) and from the MTAS measurement~\cite{Rasco_92Rb} (red).}
\label{beta-feeding}
\end{figure}
The standard GRIFFIN presorting and data-corrections procedures such as $180^{\circ}$ coincidence summing and cross-talk corrections were employed in this analysis~\cite{GRIFFIN_NIM}. %in this analysis. 
More details about the experiment, data analysis, and the final data tables will be reported in a follow up publication.\\
The results of the presented work led to a significant extension of the level scheme and provide a highly detailed picture of low-spin states in $^{92}$Sr.
%An analysis of singles, double- and triple $\gamma$-ray coincidences revealed 864 newly observed $\gamma$-ray transitions depopulating 190 levels of $^{92}$Sr.
In total, we observed the population of 190 excited states in $^{92}$Sr and 864 unique $\gamma$-ray transitions. % from analysis of $\gamma$-ray singles, $\gamma$-$\gamma$ and $\gamma$-$\gamma$-$\gamma$ coincidence data.
%Spin assignments were obtained for 40 levels from $\gamma$-$\gamma$ angular correlations with only one level below 3~MeV unassigned, a further 50 levels 
In this analysis, $\gamma$-ray and $\beta$-feeding intensities were measured relative to the $\gamma$-ray decay of 815-keV $2_1^+\rightarrow0_1^+$ $\gamma$-ray transition.
The normalization to absolute $\gamma$-ray and $\beta$-feeding intensities was calculated with the assumption that the absolute $\gamma$-ray branching of the $2_1^+\rightarrow0_1^+$ transition in $^{92}$Sr per $^{92}$Rb decay of 0.032(4), measured in a dedicated experiment~\cite{BR_815Gamma_92Rb}, is correct.
%Absolute $I_{\beta}$ intensities were not directly measured in this work, but are calculated with the assumption that the absolute $\gamma$-ray branching of the $2_1^+\rightarrow0_1^+$ transition in $^{92}$Sr per $^{92}$Rb decay of 0.032(4), measured in a dedicated experiment~\cite{BR_815Gamma_92Rb}, is correct. 
From this, we extract a ground-state to ground-state $\beta$ feeding of 94.6(8)$\%$, which is in reasonable agreement with 91(3)$\%$ measured in the aforementioned MTAS study~\cite{Rasco_92Rb}, but not with the 87.5(25)$\%$ measured in an earlier TAS study~\cite{Zakari_92Rb}.
\\
The individual $\beta$-feeding probabilities obtained from this work are compared to the previous MTAS study ~\cite{Rasco_92Rb}, and presented at the top of Fig.~\ref{beta-feeding}.
We elect to compare our results to Ref.~\cite{Rasco_92Rb} rather than Ref.~\cite{Zakari_92Rb} as the analysis of the former was performed with a greater level of statistics and utilized a less model dependent analysis. 
A one-to-one comparison of individual $\beta$~branches is only possible below 3~MeV, where the MTAS analysis employed discrete nuclear levels, while above 3~MeV the MTAS data measures $\beta$~feeding in 25-keV wide energy bins.
The $\beta$-feeding probabilities for levels below 4.5~MeV are in excellent agreement between the methods, while at higher energies, the $\beta$-feeding structure from the MTAS data maps very well with the highly detailed $\beta$~feeding results from this analysis.
Given the significant difference in sensitivity between these two methods, a more robust comparison, particularly the effect of Pandemonium, is to examine the cumulative $\beta$-feeding to excited levels in $^{92}$Sr as shown in the bottom of Fig.~\ref{beta-feeding}.
%The cumulative $\beta$-feeding to excited levels in $^{92}$Sr obtained from this analysis is compared to results from the MTAS study~\cite{Rasco_92Rb} in Fig.~\ref{beta-feeding}.
%Here we observe excellent agreement between these two methods which is 
The level of consistency between the results of these two methods is remarkable given the number of both levels populated in $^{92}$Sr and $\gamma$-ray transitions between these levels and provide a high degree of confidence in the final log~$ft$  obtained in this work.
Although the MTAS results obtain a larger $\beta$-feeding to levels between 4.5 and 6~MeV, the results are within measurement uncertainties. 
Additionally, the consistency of $\beta$-feeding below 4.5~MeV between the methods indicates that the indirect population of lower-lying levels from higher-lying levels has been accounted for thoroughly.
This highlights the capabilities of modern HPGe detector arrays in suppressing the Pandemonium effect to a significant degree for decays with low $\gamma$-ray multiplicity, even in cases with large $Q_{\beta}$ windows with highly fragmented $\beta$-intensity distributions. 
The 6030.0-keV level in $^{92}$Sr is a prime example of this Pandemonium effect.
Prior to this work, only three $\gamma$-ray decays depopulating this level were known.
However this can now be extended to 40 $\gamma$-ray decays with $\sim10\%$ of the total $\gamma$-decay strength coming from 23 branches, each with less than $1\%$ intensity. 
Without the large statistics from these data combined with the resolving power of GRIFFIN, such weak branches would be near impossible to identify, leading to misplacement of $\beta$-feeding intensities i.e. the Pandemonium effect. 
This work highlights the superior performance of GRIFFIN with the respect to previous efforts comparing $\beta$-feeding intensities of $^{97}$Ag~\cite{HRS-TAS-97Ag} and $^{150}$Ho~\cite{Algora-150Dy}
from TAS data and high-resolution data collected with the CLUSTER CUBE setup~\cite{EUROBALL}, accessing higher-energy $\gamma$-ray transitions and final levels.
\\
%\begin{figure}[t]
%\includegraphics[width=\columnwidth]{cumulativebetafeedingletter.pdf}
%\caption{Cumulative $\beta$-feeding intensity in $^{92}$Sr excluding the ground-state decay measured in this work (blue) and from the MTAS measurement~\cite{Rasco_92Rb} (red).}
%\label{beta-feeding}
%\end{figure}
%The level scheme of $^{92}$Sr obtained from this analysis is presented in Fig.~\ref{level-scheme}.
\begin{figure}[t]
\includegraphics[width=\columnwidth]{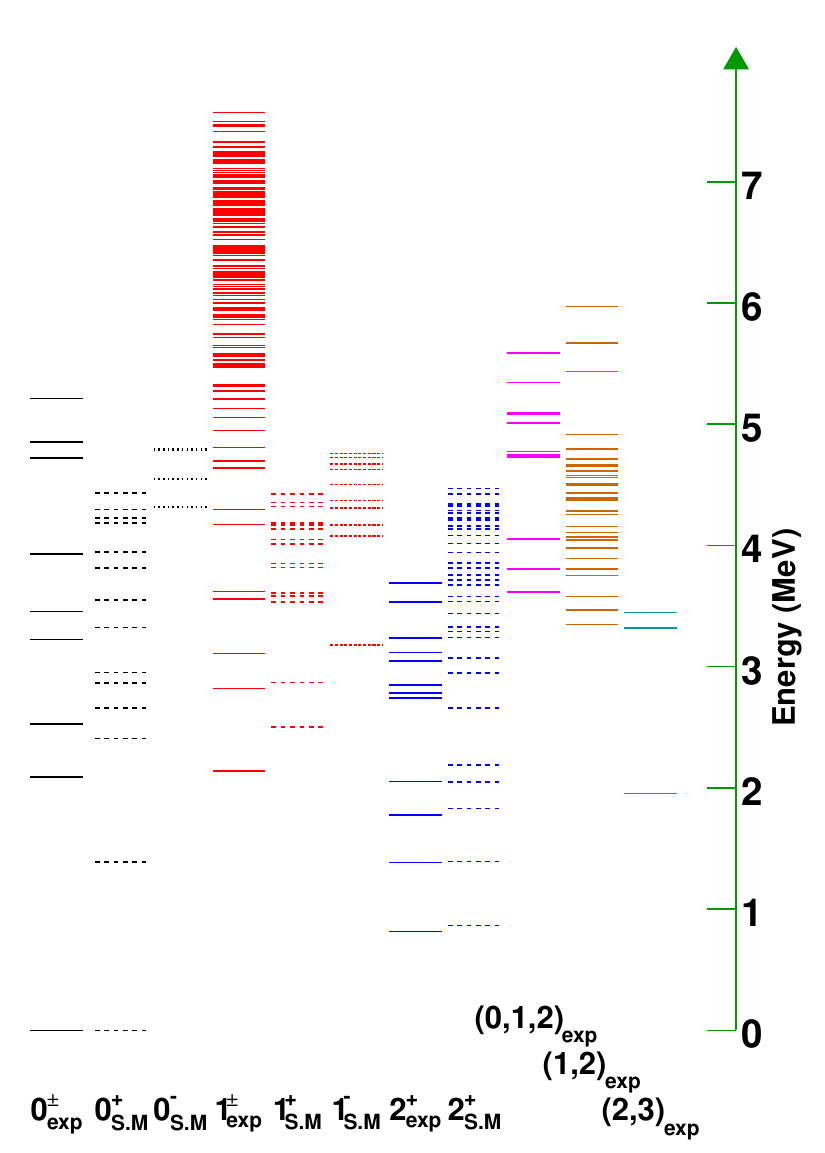}
\caption{Level scheme of low-spin states in $^{92}$Sr populated via $\beta^-$ decay of $^{92}$Rb. Experimental levels are drawn as solid lines for $J=0$ (black), $J=1$ (red) and $J=2$ (blue), while levels resulting from shell model calculations are presented with short dashes (positive parity) and long dashes (negative parity).}
\label{level-scheme}
\end{figure}
\begin{figure*}[]
\includegraphics[width=\textwidth]{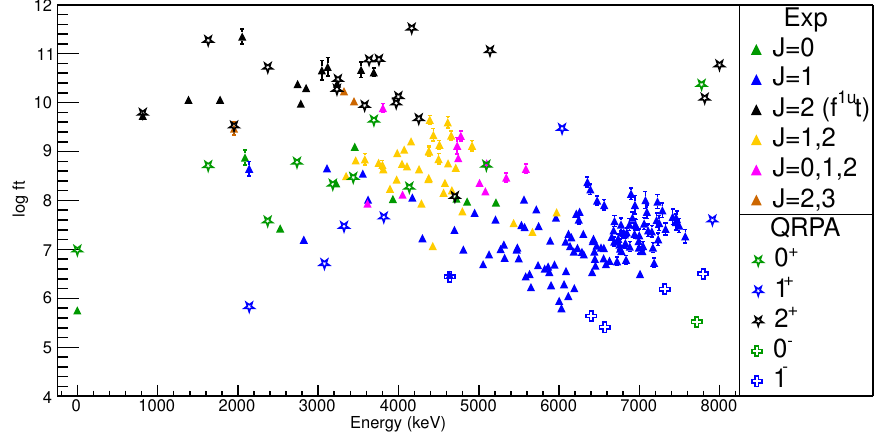}
%\includegraphics[scale=0.8]{log-ft-plot-colour-scaling.pdf}
%\includegraphics[scale=0.8]{log-ft-92Sr-Exp-QRPA.pdf}
%\caption{log~$ft$ values for the $\beta$-decay of $^{92}$Rb to final levels in $^{92}$Sr separated by spin with $J=0^+$ (green), $J=0,1,2$ (magenta), $J=2,3$ (brown), $J=2^+$ (black) are unique first-forbidden decays, $J=1$ (blue) and $J=1,2$ (yellow to red) are presented with a color range representing the strength of the absolute $\gamma$-ray branching ratio to the ground state of $^{92}$Sr.}
\caption{Experimental log~$ft$ values for the $\beta^{-}$~decay of $^{92}$Rb to final levels in $^{92}$Sr separated by spin, with $J=0^+$ (green), $J=1$ (blue), $J=1,2$ (yellow), $J=0,1,2$ (magenta), $J=2,3$ (brown) and $J=2^+$ (black) are unique first-forbidden decays (f$^{1u}$t). MCM calculated log~$ft$ values values for the $\beta^{-}$~decay of $^{92}$Rb to final levels in $^{92}$Sr separated by spin with $J=0^+$ (green star), $J=0^-$ (green cross), $J=1^+$ (blue star), $J=1^-$ (blue cross) and $J=2^+$ (black star).}
\label{log-ft-plot}
\end{figure*}
The $\beta^{-}$~decay of $^{92}$Rb, with its $0^{-}$ ground state, may populate $0^{-}$ and $1^{-}$ levels in $^{92}$Sr via allowed decays, while first-forbidden decays populate $0^{+}$, $1^{+}$, and $2^{+}$ levels.
%and $2^{+}$ levels in $^{92}$Sr via allowed and first-forbidden decays.
Spin assignments were obtained for 55 levels from analysis of $\gamma$-$\gamma$ angular correlations following ``Method 3" outlined in Ref.~\cite{SMITH201947} leaving only one level below 3~MeV unassigned.
The spins of a further 30 levels were constrained based upon log~$ft$ values. 
Here, a $J^{\pi}$=2$^+$ assignment for a level is excluded if the log~$ft$ value for a unique first-forbidden decay, is less than 8.5.
This condition is applied following the conclusions from a recent systematic study of log~$ft$ values~\cite{TURKAT2023101584}. %of log~$ft$ values
Levels which meet this requirement and exhibit $\gamma$-ray decays to $0^+$ states can therefore, only possess a $J=1$ spin.
Finally, levels above 5~MeV with large $\gamma$-decay branches to the ground state are tentatively assigned $J=1$ spin. 
This assumption is based on the dominance of the $E1$ and $M1$ gamma strength functions (gSF) over the $E2$ gSF~\cite{gSF-ref}.\\
% as, at high excitation energies, $E2$ decays compete weakly against available $M1$ and $E1$ decays in accordance with the $\gamma$ Strength Function.
%Finally, levels above 5~MeV with large $\gamma$-decay branches to the ground state are tentatively assigned $J=1$ spin as $E2$ decays compete weakly against $M1$ and $E1$ decays for high-energy $\gamma$-ray transitions in accordance with the $\gamma$ Strength Function.
%This is reflected in the multipolarity ($L1+L2$) mixing ratios extracted from $\gamma$-$\gamma$ angular correlations, where of the 23 measured $\gamma$-decays from levels above 4~MeV, the $L=2$ component of each transition is below $1\%$. \\
The level scheme of $^{92}$Sr obtained from this analysis is compared to large-scale Shell-Model (LSSM) calculations in Fig.~\ref{level-scheme}.
These calculations were performed using the KSHELL software~\cite{KSHELL} with the $jj45pnb$~\cite{jj45pnb} Hamiltonian and using standard effective charges and bare g-factors.
Starting with a $^{78}$Ni ($Z=28$, $N=50$) closed core, the single-particle model space for the calculations consists of the proton orbitals $1f_{5/2}$, $2p_{3/2}$, $2p_{1/2}$, and $1g_{9/2}$, and the neutron orbitals $1g_{7/2}$, $2d_{5/2}$, $2d_{3/2}$, $3s_{1/2}$, and $1h_{11/2}$. 
The $1h_{11/2}$ is truncated to permit a maximum occupancy of six neutrons due to computational reasons.
The LSSM calculations obtain an good reproduction of the experimental levels up to 5~MeV excitation energy despite the complexity of the nuclear level scheme and ambiguous $J^{\pi}$ assignments of many levels situated above 3.5~MeV.\\
The microscopic structure of the $\beta^{-}$~decay of $^{92}$Rb is further revealed by investigating the log~$ft$ values of transitions to final states in $^{92}$Sr, which are presented in a novel way as a function of level energy in Fig.~\ref{log-ft-plot} exhibiting an unparalleled level of detail.
The figure demonstrates the evolution of log~$ft$ values over the entire energy range of the $Q_{\beta}$ window and reflect the evolving overlap of the wavefunctions between the ground state of $^{92}$Rb and the final states in $^{92}$Sr.
Here, we observe a gradual change from almost elusively first-forbidden transitions populating lower-lying levels to predominantly allowed transitions populating high-lying $J=1^-$ levels in the PDR region. \\
%Excited states in $^{92}$Sr below 4~MeV are populated almost elusively via first-forbidden transitions. 
%However, almost all levels in $^{92}$Sr populated above 5~MeV are either $1^+$ or $1^-$.
These results are accompanied by calculated log~$ft$ values using the Multiple-Commutator Model (MCM) to reveal the underlying structure of these levels.
The MCM, was introduced in Ref.~\cite{Suhonen1993} for one-phonon final states and further extended in Ref.~\cite{Civitarese1994} to include also two-phonon final states. 
The formalism of first-forbidden unique (decay transitions from the $0^-$ ground state of $^{92}$Rb to the $2^+$ states in $^{92}$Sr) and non-unique ($0^-\to 0^+,1^+$) $\beta$ decay, used in this work, was introduced in Ref.~\cite{Suhonen1993}. 
The allowed Fermi and Gamow-Teller transitions ($0^-\to 0^-,1^-$) are thoroughly described in Ref.~\cite{Suhonen2007}.
%UPDATED VERSION FROM JUONI
The MCM is a higher-RPA (Random-Phase Approximation) approach using the QRPA (Quasiparticle-RPA) to compute the wavefunctions of states in the double-even reference nucleus $^{92}$Sr and the pnQRPA (proton-neutron QRPA) to compute the wave function of the $0^-$ ground state of the double-odd $^{92}$Rb nucleus.
MCM is based on an effective Hamiltonian consisting of a single-particle mean field based on Woods-Saxon single particle energies~\cite{Suhonen2007} and on an effective interaction derived from the one-boson-exchange G-matrix of the Bonn type~\cite{Holinde1981}.
A large no-core single-particle basis consisting of 22 proton and neutron single-particle states from the $0s$ to the $0h_{9/2}-1f-2p-0i_{13/2}$ major shell, was used in the calculations.
The MCM approach is based on BCS quasiparticles~\cite{Suhonen2007} created by scaling the pairing part of the G matrix (see \cite{Suhonen1993,Civitarese1994}) by the pairing strengths $g_{\rm pair}^{\rm (p)}=0.8698$ for protons and $g_{\rm pair}^{\rm (n)}=0.8566$ for neutrons in order to reproduce the phenomenological proton and neutron pairing gaps, as described in \cite{Suhonen2007,Suhonen1988a,Suhonen1988b}.\\
The computation of the wave function of the $0^-$ ground state of $^{92}$Rb features two adjustable strength parameters of the adopted G-matrix based Hamiltonian, namely the particle-hole parameter $g_{\rm ph}$ and particle-particle parameter $g_{\rm pp}$ \cite{Suhonen1988a,Suhonen1988b}. 
The adopted values for these parameters were $g_{\rm ph}=0.9722$ and $g_{\rm pp}=0.4577$. 
The wavefunctions of the ground state and excited one-phonon and two-phonon states in $^{92}$Sr were computed using the QRPA. 
%The two-phonon states were constructed from two one-phonon states according to the method developed in \cite{Civitarese1994}. 
The two-phonon states were constructed from two one-phonon states and the transition amplitudes were derived according to the method developed in \cite{Civitarese1994}. 
The wavefunctions of the one-phonon multipole states $J^{\pi}$ of QRPA were created by scaling the corresponding particle-hole parameter $G_{\rm ph}$ and particle-particle parameter $G_{\rm pp}$ of the effective G-matrix interaction separately for each multipole.
The corresponding parameter values are shown in the table~\ref{myQRPATable}.\\
\begin{table}[t]
\label{myQRPATable}
\centering
\caption{QRPA parameters for the particle-hole, $G_{ph}$, and particle-particle, $G_{pp}$, of the effective G-matrix interactions.}
\begin{tabular}{|c|ccccc|}
%\begin{tabular}{|P{1cm}|P{1.25cm}P{1.25cm}P{1.25cm}P{1.25cm}P{1.25cm}|}
\hline
$J^\pi$ & $0^-$ & $1^-$ & $0^+$ & $1^+$ & $2^+$ \\
\hline
$G_{\rm ph}$ & 1.0000 & 0.5291 & 0.3729 & 1.0125 & 0.5115 \\
$G_{\rm pp}$ & 1.0000 & 2.0766 & 0.5724 & 3.9100 & 1.0470 \\
\hline
\end{tabular}
\end{table}
From our experimental data, we observe sizable $\beta$~feeding to a large number of $1^-$ levels in $^{92}$Sr which, with the exception of the ground-state decay, exhaust almost all of the observed $\beta$~strength.
At the same time, the MCM calculations produce four $1^-$ levels above 6~MeV corresponding to $1p$-$1h$ configurations. 
%The calculated log~$ft$ values are independent of the phase-space factor, therefore independent of the $Q$ value and reflect the overlap of the wavefunctions between the initial and final states.
These $1p$-$1h$ configurations become fragmented over a large number of $1^-$ levels via coupling to higher-order quasiparticle states.
This results in the $\beta$-feeding being distributed to a large number of states as observed in this work.\\
These results mirrors recent observations from the neighboring $\beta^{-}$~decay of $^{96}$Y~($J^{\pi}$=$0^-$)~$\rightarrow$~$^{96}$Zr~\cite{mashtakov_96Zr}. 
\begin{figure}[b]
\includegraphics[width=\columnwidth]{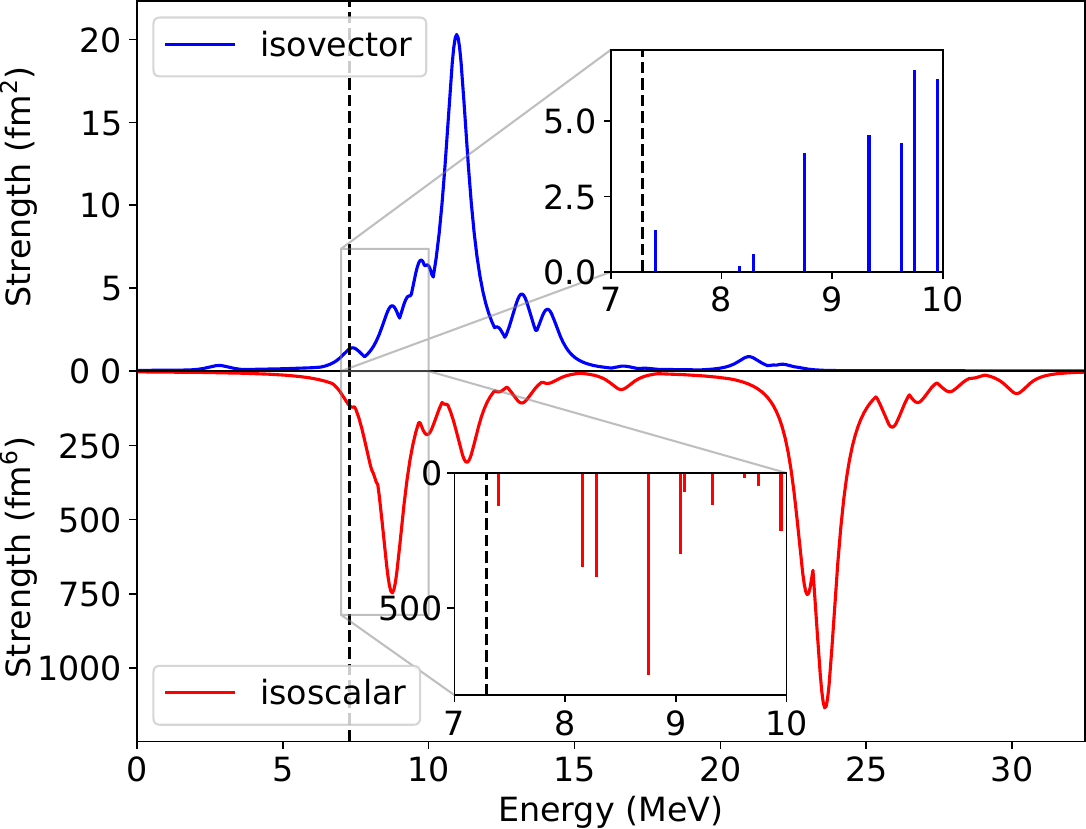}
\caption{Isovector (blue) and isoscalar (red) dipole strengths for $^{92}$Sr from spherical QRPA calculations as performed in Ref.~\cite{QRPA-PDR-92Sr}. The neutron separation energy of $^{92}$Sr is shown with a dashed line.}
\label{92Sr-pdr}
\end{figure}
Both decays are key contributors to high-energy reactor neutrinos due to first-forbidden decays to the ground-state with uniquely small log~$ft$ values and at the same time exhibit considerable $\beta$~feeding to high-lying $1^-$ levels in the PDR region of the daughter nuclei.
However, in this work, we observed the population of more than 150 levels above 4~MeV in $^{92}$Sr compared to only 60 levels in $^{96}$Zr. 
Additionally, we observed, to a far greater degree, a significantly more fragmented $\gamma$-ray decay behavior.
This is likely a consequence of the greater resolving power of GRIFFIN and relatively small background compared to the more limited experimental conditions of Ref.~\cite{mashtakov_96Zr}.
%In $^{96}$Zr, these $1^-$ levels are associated with a resonance-like structure situated at the low-energy tail of the Giant Dipole Resonance known as the Pygmy Dipole Resonance (PDR).
The $\beta$-decay data of $^{96}$Y were complemented with Nuclear Resonance Fluorescence (NRF) data and revealed that almost all levels observed in the NRF data were also populated via allowed Gamow-Teller transitions in $\beta$~decay.
The experimental data was complemented with theoretical interpretations using the Quasi-particle Phonon Model~\cite{soloviev2020theory} which revealed that the components of the $1^-$ PDR levels populated via $\beta$~decay contribute with small amplitudes to their complex wavefunction with the $\beta$-decay process more likely to populate more complex $2p-2h$ configurations.
For $^{92}$Sr, the isovector and isoscalar dipole strength is calculated using spherical QRPA calculations and presented in Fig.~\ref{92Sr-pdr}.
These calculations employed the same approach that was adopted for the neighboring nuclei $^{90-94}$Zr and $^{92-100}$Mo~\cite{QRPA-PDR-92Sr}.
The calculations reveal weak electric dipole strength situated near the neutron separation energy carried by the $1^-$ states populated via the $\beta^{-}$~decay of $^{92}$Rb. 
%The wavefunction of the lowest-lying isoscalar strength at 7.4~MeV, shown in the inset of Fig.~\ref{92Sr-pdr} is dominated by a single-particle $\nu d_{5/2}$ to $\nu f_{7/2}$ transition.
The wavefunction of the lowest-lying isoscalar strength at 7.4~MeV, shown in the inset of Fig.~\ref{92Sr-pdr} is dominated by a single-particle $(\nu d_{5/2})^{-1}$ to $\nu f_{7/2}$ transition.
However, the wavefunction of isovector electric dipole strength situated close in energy is comprised of multiple proton and neutron excitations.
%the isoscalar peak at 7.4 MeV has as major single-particle transition the neutron n1d52 - n1f7/2 transition (with amplitude 0.87) and the corresponding isovector strength has been distributed to many components.
These results, combined with the previous study of $^{96}$Y decay, indicate the PDR plays a role in the $\beta$-decay spectrum of fission fragments, which posses low-spin, negative-parity ground states and large $Q_{\beta}$ values.
Consequently, a detailed decay study of $^{142}$Cs, the third main contributor to high-energy reactor neutrinos is an ideal candidate to further investigate this phenomenon. \\
In summary, we have studied the $\beta^{-}$~decay of $^{92}$Rb, one of the most important decays for the RAA with the GRIFFIN spectrometer at TRIUMF.
Thanks to the capabilities of GRIFFIN, the fine structure of this decay is revealed in unprecedented detail.
%The $\beta$-feeding probabilities measured in this work are in excellent agreement with earlier MTAS results, demonstrating the capabilities of modern HPGe detector arrays in suppressing the Pandemonium effect and at the same time to investigate the fine structure of these decays in unparalleled detail.
We observe sizable $\beta$~feeding to $1^-$ levels in the energy region of the Pygmy Dipole Resonance in $^{92}$Sr, mirroring recent observations of $^{96}$Zr but with many more levels populated and a more fragmented $\gamma$-ray decay behavior. 
These results provide the most detailed picture of $1^-$ levels of the PDR in a nucleus far from stability and will serve any future Coulomb Excitation study of $^{92}$Sr with vital spectroscopic data  to extract the full electric dipole strength.
The Multiple-Commutator Model calculations predict several $1p-1h$ configurations in this energy region being responsible in populating these levels with intensities comparable to the population of low-lying levels which are significantly favored by the effective $Q$ value. 
%These results highlight the dependence of reactor antineutrino spectra on the nuclear structure of fission fragments themselves.
%These results highlight the role of the fission fragments nuclear structure that shape the reactor antineutrino spectrum.
These results highlight the role of the nuclear structure of fission fragments in shaping the reactor antineutrino spectrum.
Finally, the $\beta$-feeding probabilities measured in this work compare very favorably with the most recent MTAS study and demonstrate the ability of state-of-the-art HPGe detector arrays to suppress the Pandemonium effect and provide highly complementary data to TAS studies for reactor physics.
\\
%\section{Acknowledgements\label{sec:Acknowledgements}}

\textit{Acknowledgements}$-$This work would not have been possible without the efforts of the operations and beam delivery staff at TRIUMF for providing the radioactive beam.
This work was supported in part by the Natural Sciences and Engineering Research Council of Canada and by the U.S. Department of Energy, Office of Science, Office of Nuclear Physics, under contract number DE-AC02-06CH11357.
The infrastructure of GRIFFIN has been funded through contributions from the Canada Foundation for Innovation, TRIUMF, Simon Fraser University, University of Guelph, British Columbia Knowledge Development Fund, and the Ontario Ministry of Research and Innovation. 
TRIUMF receives funding through a contribution agreement through the National Research Council Canada.
M. Ramalho acknowledges support provided by the Oskar Huttunen Foundation.
The authors wish to acknowledge CSC – IT Center for Science, Finland, for generous computational resources.
\bibliographystyle{apsrev4-1}
\bibliography{sample}
\end{document}